# A parallel algorithm for Delaunay triangulation of moving points on the plane

*Nazanin Hadiniya, Mohammad Ghodsi*

*Computer Engineering Department*

*Sharif University of Technology*

*Tehran, Iran*

*Abstract*-- Delaunay Triangulation( DT) is one of the important geometric problems that is used in various branches of knowledge such as computer vision, terrain modeling, spatial clustering and networking. Kinetic data structures has become very important in computational geometry for dealing with moving objects. However, when dealing with moving points, maintaining a dynamically changing Delaunay triangulation can be challenging. So, In this case, we have to update triangulation repeatedly. If the points move so far, it's better to rebuild the triangulation. One approach to handle moving points is to use an incremental algorithm. For the case that points move slowly, we can give a faster algorithm than rebuilding. Furthermore, sequential algorithms can be computationally expensive for large datasets. So one way to compute as fast as possible is parallelism. In this paper, we propose a parallel algorithm for moving points. we propose an algorithm that divides datasets into equal partitions and give every partition to one block. Each block satisfay the Delaunay constraints after each time step and uses delete and insert algorithms to do this. We show this algorithm works faster than serial algorithms.

*Keyword*-- *Delaunay triangulation, parallel, moving points, kinetic*

## I. INTRODUCTION

Delaunay Triangulaton is a fundamental problem in geometric and plays an important rule in terrain modeling, scientific data visualization, surface construction, and so on [1]( figure 1). A DT of the dataset is a triangulation that satisfies several conditions. The most important condition is that the circumcircle of a triangle includes only the vertices of it. This property ensures that the triangles do not overlap or intersect, which makes them ideal for various geometric algorithms and computations. One of the other properties of DT is that every triangle's minimum angle is as large as possible, so it prevents DT from having long narrow triangles. The other aspect of this triangulation is that it is unique for a given dataset [2]. Every DT has a dual graph called a Voronoi Diagram, which is formed by using the circumcenters of each delaunay triangle. This property makes Delaunay triangulation useful for applications such as finding the boundary of a dataset. The DT can also be extended to higher dimensions, such as generating tetrahedral meshes in 3D.

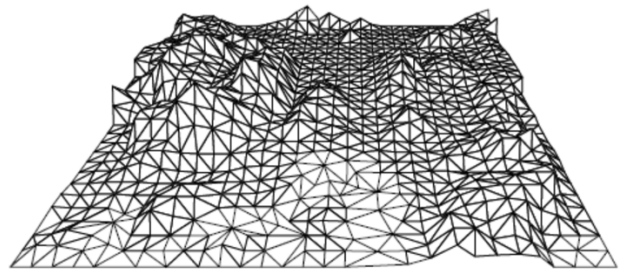

*Figure 1: the usage of DT in terrain modeling*

Larg datasets encourage programmers to develop parallel algorithms in order to calculate DT in a faster time. Many parallel algorithms were developed for fixed points DT and most of them used divide and conquer methods. This study focuses on the parallelism of DT for larg moving datasets. The proposed algorithm receives as input a set of points with their x and y Coordinates. But the points location vary during the time and the triangulation should be updated. Different parameters like how far points move, effect the algorithm. If the points move so far, it's better to rebuild the triangulation but for the limited range of movement of the points we can give more efficient algorithms.

Kinetic DT is a computational technique used in computational geometry, primarily for dynamic geometric environments. It is particularly useful for scenarios where the triangulation needs to be continuously updated as points move or change positions over time. It is also very important for different applications that involves motion planning, such as video games, virtual reality, dynamic simulation and robotics [3]. The other use of kinetic DT is that it can be used in Geographic Information Systems(GIS) applications for dynamic map updates, location-based services, and network analysis. Overall, Kinetic DT helps to efficiently track and adapt the triangulation structure in dynamic environments, enabling various applications that require real-time updates and accurate geometric calculations.

The proposed algorithm is explored in terms of the time complexity and comparison with serial form of it, as explained in Section 4. The most important features and constraints of the proposed algorithm are as follows:

- It reconstructs DT in regular time steps by using insert and deletion approach.
- It can handle large datasets.
- We consider that in each time step just one point in every block moves. This can make the relation between blocks lower.
- Every point can just move to its right, left, up or down direction.

The rest of the paper is organized as follows:

sections 2 provides an overview of existing algorithms in the field of kinetic DT. Section 3 explains the proposed algorithm and Section 4 presents analysis and results of comparing the proposed algorithm with the serial form of the algorithm. We also show time complexity and speed up of the proposed algorithm. Section 5 gives the summary and conclusion of the paper.

## II. RELATED WORKS

The importance of DT in different fields has encouraged many researchers to develop algorithms for its various aspects. For example, there are many algorithm for the simple form of DT, the parallel form of it and the usage of it. There have been also several papers for the sequential and parallel form of the DT with moving points or kinetic DT, as discussed next.

De Berg et al. [4] study kinetic convex hull and DT in the black box model under the assumption on the point movement and time steps to optain provably efficient solutions. In every time step, every point just move a limited distance $d_{max}$. the value of $d_{max}$ shows that any point p in the point set P, the disk of radius $d_{max}$ contains at most k points for a given constant k. they describe two algorithms for recomputing DT at time t using DT of time t-1: the first one is move and flip approach and the second one is insert and delete approach. In the first approach the running time is $O(k^2\Delta^2_k \log n)$ and for the second one the running time is $O(k^2\Delta^2_k)$.

Devillers et al. [5] worked on the two-dimensional DT and reviewed different methods to compute DT of a set of moving points. These methods classified in kinetic methods, methods using checking and points insertion and deletion and weak Delaunay approaches. The first method looks between timestamps, the second only focus on timestamps and the third one goes one step forward and may forget some modifications so this one uses for approximated DT.

De castro et al. [6] introduced the concepts of tolerance and safe region of a vertex, such that as long as the point stay in its tolerance, it doesn't need to be checked. For two-dimensional, their algorithm is up to six times faster than rebuilding. But for three-dimensional, rebuilding is faster than their algorithm. Heinich et al. [7] studied the problem of maintaining the DT of moving two-dimensional points on the GPU with discrete time steps. they showed that their structure is useful for answering proximity queries and they published a paper [8] for fixed-radius nearest neighbors problem.

Yoo et al. [9] proposed an algorithm for two-dimensional DT of moving points. They assign every point to a processor which they can only communicate with the points that has an edge with them in DT. Because points move, the topology of processors change after eache time step but this is done automatically by local operations without any global data.

Schaller et al. [10] describe algorithms for three dimensional kinetic DT. To do this, they used a three-simplex data structure. The data structure is obtained by adding neighborship entries to every simplex and sorting the tetrahedra within a list. They also describe a simple incremental algorithm for deletion of points when their location change.

Kaplan et al. [3] proposed a simple randomized scheme for kinetic DT in the plane. They consider that points move continuously along piecewise algebraic trajectories of constant description complexity.

Dinas et al. [11] proposed an algorithm for kinetic DT in order to use it for modeling collision detection. To do this, they proposed a Kinetic Constrained Delaunay Triangulation, which is as Delaunay as possible and represents obstacles by constrained edges and objects by vertices. Second, they contribute with this model that guarantees both constrained edges and the Delaunay Triangulation structure. They compare two different models: the Kinetic Constrained Delaunay Triangulation (KCDT) model and The Kinetic Delaunay Constrained Triangulation (KDCT) model. At the end, they reached to this conclusion that in most cases the KCDT is slightly better than KDCT.

### III. PRELIMINARIES

In this section we introduce some notation and definitions for solving our problem. First we talk about the structure we use to divide dataset into equal number of points for each block. Then we talk about some notation of the algorithm. After these definitions, in the next section we represent our algorithm.

**Data structure**

To do parallel computation, first we should know how to divide the point set and tasks into equal partitions. So that every block has approximately the same amount of work and data. To do this, we use a structure like quad-tree structure for our dataset( figure 2). In quad-tree structure every node of the tree has four children. First we find the minimum square that covers all the points and consider a threshold for the points that each square should has in it. Then this square divides into four equal squares. Each of these squares divides into four equal squares. This division continues until all squares have equal or less than the threshold number of points in them. In our structure every block is connected to four other blocks around it. For this, we can consider that every block has an connection with its left, right, up and down block. These connections are useful for our algorithm so that when points move to other neighbor they can communicate easyily(figure 3).

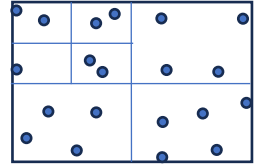

*Figure 2: part of a quad-tree structure with 2 as its threshold*

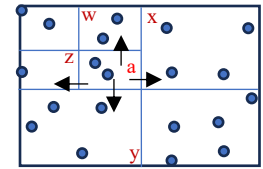

*Figure 3: neighbors of block a: w, x, y and z.*

**Notation and definitions**

In this part we want to define all the variables that we use for the algorithm. As mentioned in [4] when points move so far in each time step, rebuilding the triangulation is better than using different methods for updating the triangulation. So we consider that at each time step not so many points move and they move only a limited distance. We name this limitation d in our algorithm. To find a proper value for d we do like what paper [4] have done. We get a constant number N as an input. Then we find the Nth nearest neighbor of each node. The minimum distance among all these Nth nearest neighbor is the value of d.

For $\forall p \in P$ and $\forall n \in neighbor\ p$, calculate distance(pn) so we could find Nth nearest neighbor of each node and save it in $dist_p$. Then we use mindistance(P) to find the minimum value among all the Nth nearest neighbor of points.

$$\text{mindistance}(P) = \min_{p \in P} dist_p$$

after each time step we should update the value of d in such a way that it remains the minimun value of mindistance(P) during the time:

$$d = \min_{t} mindistance(P_t)$$

time steps should be long enough that we could update the triangulation. After each time step the Delaunay conditions are satisfied but during updating it might not be satisfied. So, only the DT after each time step is completely valid.

In our algorithm we consider that only one point move in each time step in each block. So, for each block we consider a variable which shows the moved point. We name this list for each block i as $movepoint_i$.

$$movepoint_i = p \in P_i \text{ such that}$$
$$location_t(p) \neq location_{t+1}(p)$$

the moved point and its new location is chosen randomly at each time step in each block. So, we have randomly moved point in each block.

The other concept that was used by De Berg [4] is N-spread. They used this concept to evaluate their algorithm. The N-spread was defined as follows:

$$\Delta_N(P) = diam\ (P)\ /\ mindistance(P)$$

In our algorithm, we compute th N-spread value locally. diam(P) is the value of diameter of our dataset P and we consider this value is fixed during the movement of the points. To compute the N-spread value, we first compute the diameter of each block. Since we have used quad-tree like structure, after each division of our dataset, the diameter of it is halved. So, the diameter of each block is in the order of O(log diam(P)).

Since the value of mindistance(P) may change during the movement of the points, the value of N-spread may change too. So, the value of N-spread is its maximum value during all the time steps of the algorithm for each block.

$$diam\ (P_i) = O(\log\ diam(P))$$
$$\Delta_N(P_i) = diam\ (P_i)\ /\ mindistance(P)$$
$$= O(\log\ diam(P))\ /\ mindistance(P)$$

IV. SOLUTION APPROACH

Now, in this section we want to represent our algorithm for solving kinetic DT. Our goal is to speed up the implementation and represent an efficient algorithm. Most solutions for this problem are in the serial form. So, we want to use parallelism in order to have a faster algorithm.

At the beginning, we use a parallel form of quad-tree to divide our data into equal partitions [12]. We modify this algorithm such that every block has at most a constant number of points. This constant number depends on the total number of points and the memory of each block. We name this sub-algorithm Quad-treeDvision.

After partitioning the data, each block has approximately the same number of points. each of them has a list of points, list of edges corresponds to the points, list of edges to the points that are in other blocks which are their neighbors and their distance. In each block the value of d calculates locally and after that the total minimum value of d calculates in parallel and broadcasts to all the blocks. we name this sub-algorithm d-calculation. This value changes during the algorithm because after moving the points, the distance between the points and their neighbors change.

At a certain time the movepoint variable of every block updates randomly and after that, all the blocks start to satisfy the Delaunay constraints locally. To do this we use two sub-algorithms, one of them for deleting moved points from their current location and the other for inserting these points in their new location. Some of these new locations don't belong to point's current block. For these points, we need communication between blocks. after it is found that the point isn't within its acceptable range, its block send the point to its new block. Then this new block insert the point into its triangulation.

figure 4 shows the main algorithm for kinetic DT.

---

**Main algorithm**

---

*// input: a dataset consists of points and edges between them that satisfies Delaunay constraints*
*// output: the DT of the dataset after each time step and moved points.*
1. t = 0
2. **While** (true)
3. {
4.    d = d-calculation(dataset)
5.    **Quad-treeDvision**(dataset, d)
6.    **foreach** block i **do** in **parallel**
7.      **Update**(movepoint$_i$)
8.      **Delete**(movepoint$_i$, dataset$_i$(t-1))
9.      **if new-block**(movepoint$_i$) == i
10.         **Insert**(movepoint$_i$, dataset(t))
11.      **else**
12.         j = **new-block**(movepoint$_i$)
13.         **Insert**(movepoint$_j$, dataset(t))
14.    **return** dataset(t)
15.    t++
16. }

*Figure 4: the main algorithm for parallel kinetic DT*

**Insertion and Deletion**

There are different algorithms to insert a point into a dataset to satisfy Delaunay constraints. One way to insert a point is to use incremental algorithm for DT. Bowyer-Watson algorithm [13] is a popular incremental algorithm that we can use. In each time step we have at most O(B) moved points that they process simultaneously in each block. Time complexity of these points depends on the potential edges to remove or insert. We will discuss more in the next section.

Deletion of the points is more complicated than insertion of them. To delete a point form its current location, there are different algorithms too. For example Schaller et al [10], Chin et al. [14] and Devillers [15] worked on the problem of deletion. If the deleted point has small degree, the Chin's algorithm is slow and because this algorithm is complicated its better to use a slower but simpler algorithm proposed in the Devillers paper.

## V. ANALYSIS

In order to evaluate our algorithm, we discuss about its time complexity in this section. A challenge in parallel algorithms is communication between prossesors. In our algorithm, according to our data structure time complexity of this communication for moving points is o(log(B)).

**lemma 1** in our algorithm, every point can only move in four directions: left, right, up and down.

**Theorem 1** we can guarantee that every point in each time step can cammunaicate with at most four other blocks in o(log(B)). in the case that one point only move in each time step, each block communicates with at most one of its neighbors.

**Proof**. According to the lemma 1, every point moves in four directions. So, if it is in the border of its block and move to the outer side of it, then it can go into one of its neighbors. According to quad-tree data structure, sending data to other blocks is in the order of $O(\log_4(B))$ because each block has four children.

□

If we could change this data structre in a way that blocks have a straight connection with their neighbors, then they don't need to move among the tree structure. as a result, its time complexity to send data is O(1).

To compute d value in each time step, we need to find the minimum value in each block. the time complexity of these communications is O(logB) too. this is because our data structure is a tree and in a tree finding minimum value is from order O(log B).

**Theorem 2** computing DT in time t from DT in time t-1 using insert and delete sub-algorithms and in quad-tree structure requires $O(N^2 \Delta_N^2(P_i) + \log B)$ time in total.

**Proof**. To update the triangulation in order to remain Delaunay, each block B executes the insert and delete sub-algorithms. For updating the position of point p, we should find the number of potential edges. Potential edges are edges which might be delete or insert because of moving points. this value has been proven in [4] which is $O(N^2 \Delta_N^2(P))$. In our algorithm, as we have discussed, the value of $\Delta_N(P)$ has changed.

This is because of division of the dataset. This value is $\Delta_N(P_i)$ for each block and since all the blocks do their work in parallel time complexity of our algorithm is $O(N^2 \Delta_N^2(P_i) + \log B)$. Furthermore, The value of $\log B$ is for the communication needed during updating the algorithm. If we could have neighbors connection, this value is O(1) and the algorithm takes $O(N^2 \Delta_N^2(P_i))$ time. □

## VI. CONCLUSIONS

In this paper, we studied the problem of DT of moving points in the plane. We propose a parallel algorithm and analyze its time complexity. To do this, the dataset divided into equal partitions by a structure like quad-tree. Then we consider that in each partition only one point moves in one of the four directions: left, right, up and down. After each time step, we use insert and delete sub-algorithms to satisfy Delaunay constraints in parallel to remain the dataset in the form of DT. This algorithm takes $O(N^2 \Delta_N^2(P_i) + \log B)$ time. We can make this algorithm better by adding connection between neighbors of the blocks so that it takes $O(N^2 \Delta_N^2(P_i))$ time and the speed up of our algorithm is $\frac{diam^2}{(\log diam)^2}$.